\documentclass[aps,pra,twocolumn,superscriptaddress,letterpaper,10pt]{revtex4-1}

\usepackage[colorlinks=true,citecolor=blue]{hyperref}
\usepackage{mathrsfs,graphicx,amssymb}
\usepackage[fleqn]{amsmath}

\newcommand{\pct}{{\footnotesize\%}}

\begin{document}

\title{Reply to ``Comment on Bose-Einstein condensation with a finite number of particles in a power-law trap''}

\author{A. Jaouadi}
\affiliation{Qatar Foundation, Tornado Tower, Floor 5, PO Box: 5825, Doha, Qatar}

\author{M. Telmini}
\affiliation{LSAMA, Department of Physics, Faculty of Science of Tunis, University of Tunis El Manar, 2092 Tunis, Tunisia}

\author{Eric Charron}
\affiliation{Institut des Sciences Mol\'eculaires d'Orsay (ISMO), UMR8214 CNRS, Universit\'e Paris-Sud, 91405 Orsay Cedex, France}

\date{\today}

\begin{abstract}
In this reply we show that the criticisms raised by J. Noronha are based on a misapplication of the model we have proposed in [A. Jaouadi, M. Telmini, E. Charron, Phys. Rev. A \textbf{83}, 023616 (2011)]. Here we explicitly discuss the range of validity of the approximations underlying our analytical model. We also show that the discrepancies pointed out for very small atom numbers and for very anisotropic traps are not surprising since these conditions exceed the range of validity of the model.
\end{abstract}

\maketitle

The study of the critical temperature $T_c$ associated with the condensation of dilute Bose gases has attracted a lot of attention in the last two decades\,\cite{CUP2008, CUP2009, Springer2013}. In this context, our work published in Physical Review A in 2011\,\cite{PRA2011} is the continuation of a previous study devoted to Bose-Einstein condensation (BEC) in blue-detuned Laguerre-Gauss optical traps\,\cite{PRA2010}. Two years after these theoretical investigations, BEC was finally observed experimentally using a hollow laser beam\,\cite{PRL2013}, confirming our prediction that condensation could be reached in such an experimental configuration.

In this series of two papers, Ref.\,\cite{PRA2011} specifically focused on the influence of the external trapping potential on $T_c$ beyond the thermodynamic limit, for a finite particle number $N$. It is well known that the value of $T_c$ is influenced by the different nature of the condensation process in homogeneous \textit{vs}. inhomogeneous gases\,\cite{RMP1999,RMP2004}. In harmonic traps, Bose-Einstein condensation is a local phenomenon in the sense that, initially, it only affects the atoms located near the trap center. On the contrary, in the case of a homogeneous gas, the phase transition is dominated by long-range atomic correlations. Investigating the crossover between these two limits of homogeneous and inhomogeneous potentials was one of the motivations for our study and we used power-law traps to bridge this gap.

With this aim, we have used the local density approximation (LDA), which assumes that the gas can be considered locally as homogenous, with a local chemical potential expressed as
\begin{equation}
\mu_{local}(\mathbf{r}) = \mu - V(\mathbf{r}),
\end{equation}
where $V(\mathbf{r})$ is the external trapping potential and $\mu$ is the chemical potential at the center of the trap. In typical experimental conditions using harmonic traps, it is known that this approximation yields very accurate results\,\cite{CUP2008}. As mentioned in our original article, it is also known that this approach fails in the fully homogeneous limit corresponding to cold atoms trapped in a flat box\,\cite{RMP1999}. For this limit, obtaining quantitatively accurate descriptions requires more elaborate theoretical methods. Our study based on LDA is nevertheless interesting for two reasons. First, it is an approach from which analytical results can be derived. It therefore provides a useful first approximation, known to be valid in usual experimental configurations with harmonic traps, against which one can compare more sophisticated theories. Second, due to its simplicity, it leads to a better understanding of the crossover between the two limits mentioned previously, of homogeneous and inhomogeneous potentials. This type of insight is more difficult to gain from other theories, which do not necessarily provide analytical results.

In our calculation of $T_c$ , the thermodynamic sums over discrete quantum states are replaced, using the LDA, by integrals in phase space. This approximation, which is very common\,\cite{CUP2008}, is of a semi-classical nature. To be valid, it requires the thermal energy $k_BT_c$ to far exceed typical quantum state energy differences\,\cite{Springer2013}. In the case of an isotropic harmonic trap, this is simply written as
\begin{equation}
k_BT_c \gg \hbar\omega,
\label{LDA}
\end{equation}
where $\omega$ denotes the trap angular frequency. In addition, in this specific case the critical temperature for Bose-Einstein condensation of an ideal gas in the thermodynamic limit is given by
\begin{equation}
T_c^0 = \frac{\hbar\omega}{k_B}\left(\frac{N}{\zeta(3)}\right)^{\!\frac{1}{3}},
\label{Tc0}
\end{equation}
where $N$ denotes the total number of particles and $\zeta(s)$ is the Rieman zeta function.

The approach derived in\,\cite{PRA2011} is an approximation and it is therefore important to evaluate its domain of applicability. In the case of an isotropic harmonic trap this task is relatively trivial. It requires to make an arbitrary but reasonable choice for the range of validity of Eq.\,(\ref{LDA}). In the following, we consider that the discrete nature of the trapped states can be considered as a continuum when 
\begin{equation}
k_BT_c > 20\,\hbar\omega.
\label{LDA2}
\end{equation}
Using Eqs.\,(\ref{Tc0}) and (\ref{LDA2}), we obtain the following limit of validity for our approach
\begin{equation}
N > 20^3 \zeta(3) \simeq 9600.
\label{validity}
\end{equation}
This simple estimation shows clearly that the LDA cannot be used with very small particle numbers. This limitation is known since a long time\,\cite{RMP1999}. In an isotropic harmonic trap, one can therefore expect LDA to hold for $N \geqslant 10^4$. From this simple evaluation, one can already conclude that the results presented by J. Noronha in Fig.\,1 of his comment for values of $N$ as low as $N=100$ are very far outside the limit of validity of our model, and they are therefore necessarily misleading. One cannot draw any serious conclusion from a comparison performed in this limit of very small particle numbers.

\begin{figure}[!t]
\begin{center}
\includegraphics[width=0.99\columnwidth]{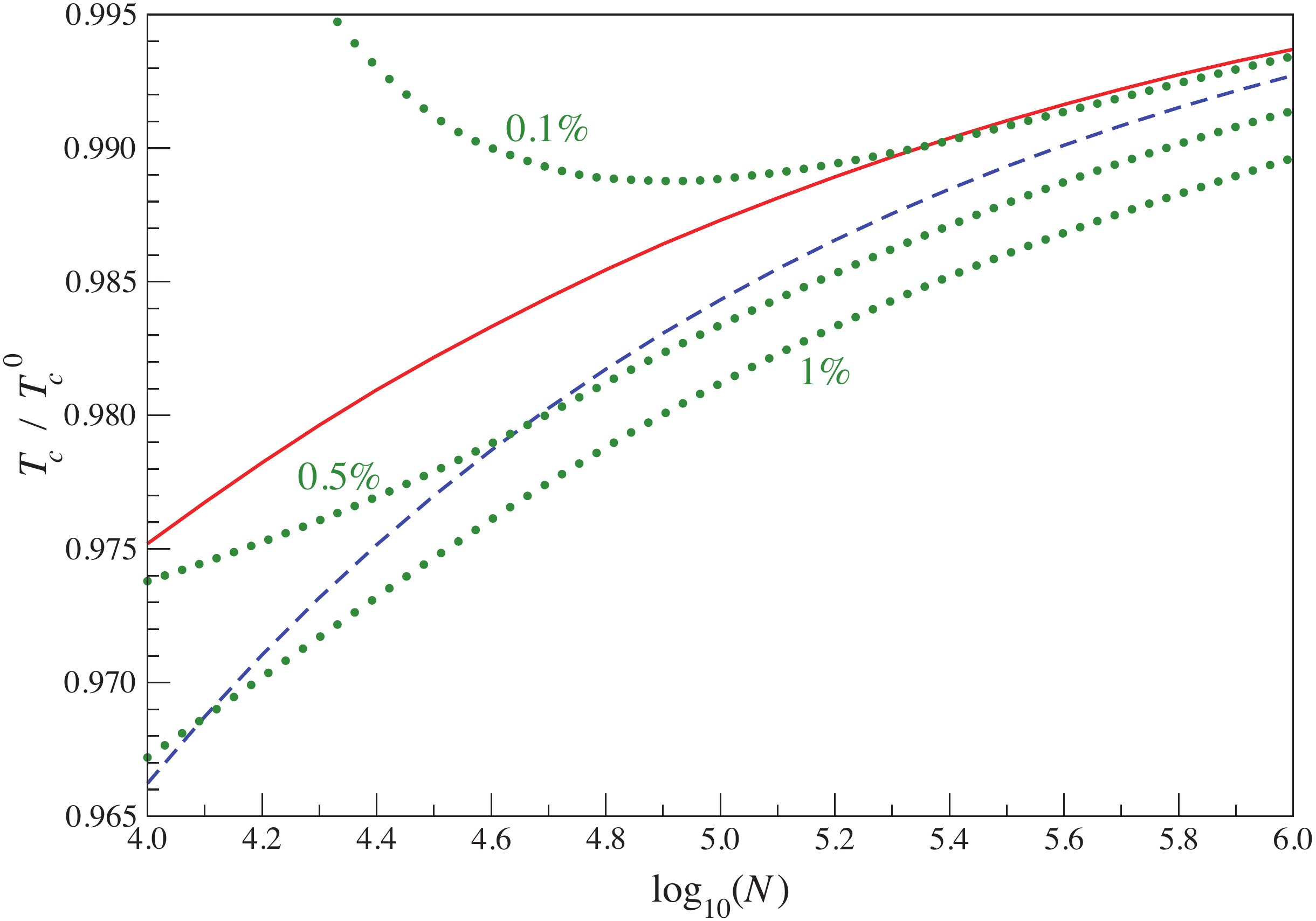}
\caption{\label{fig1}(Color online) Rescaled temperature $T_c/T_c^0$ as a function of the logarithm of the particle number. The dashed blue line shows the critical temperature calculated at first order and the solid red line shows the result of our generalized LDA formula\,\cite{PRA2011}, which includes higher-order corrections. The temperatures $T_{0.1\%}$, $T_{0.5\%}$ and $T_{1\%}$ at which the condensate fraction reaches 0.1\pct{}, 0.5\pct{} and 1\pct{} are shown in the three dotted green lines (see the three labels in the graph).}
\end{center}
\end{figure}

Now, in the case of non-harmonic traps, the limit of validity of our approximation has to be re-evaluated. In a disk-shape or in a cigar-shape harmonic trap, the inequality (\ref{LDA}) becomes
\begin{equation}
k_BT_c \gg s\,\hbar\omega,
\label{LDAs}
\end{equation}
where $s$ denotes the anisotropy parameter defined in the comment of J. Noronha. In this case, the condensation temperature in the thermodynamic limit is obtained as
\begin{equation}
T_c^0 = s^n\left[\frac{\hbar\omega}{k_B}\left(\frac{N}{\zeta(3)}\right)^{\!\frac{1}{3}}\right],
\label{Tc0s}
\end{equation}
where $n=1/3$ for a disk and $n=2/3$ for a cigar. The validity criterion (\ref{validity}) then becomes
\begin{equation}
N > \left(20\,s^{(1-n)}\right)^3 \zeta(3),
\end{equation}
or equivalently
\begin{equation}
s <\left[\frac{1}{20}\left(\frac{N}{\zeta(3)}\right)^{\!\frac{1}{3}}\right]^{\left(\frac{1}{1-n}\right)}
\label{validitys}
\end{equation}
We now apply Eq.\,(\ref{validitys}) to the two cases studied by Noronha in Fig.\,2 of his comment, with $N=10^5$. The range of validity of our approximation, as defined by Eq.\,(\ref{validitys}), is $s < 3.2$ for a disk and $s < 10.4$ for a cigar. We can clearly see in Fig.\,2 of this comment that the estimations based on our simple analytical model start to deviate from an exact numerical calculation (denoted as $T_{0.1\%}$ in this comment) when $s>3$ in the case of a disk (upper plot) and when $s>10$ in the case of a cigar (lower plot). This is in good agreement with the expected range of validity of our approximation. Clearly, one cannot draw any serious conclusion from comparisons performed in the limit of very anisotropic traps. Conversely, in the range where our approximation is justified, that is for a relatively small anisotropy, we can conclude from the calculations performed by J. Noronha that the results given by our simple analytical formula agree rather well with other more involved numerical estimations of $T_c$.

\begin{figure}[!t]
\begin{center}
\includegraphics[width=0.99\columnwidth]{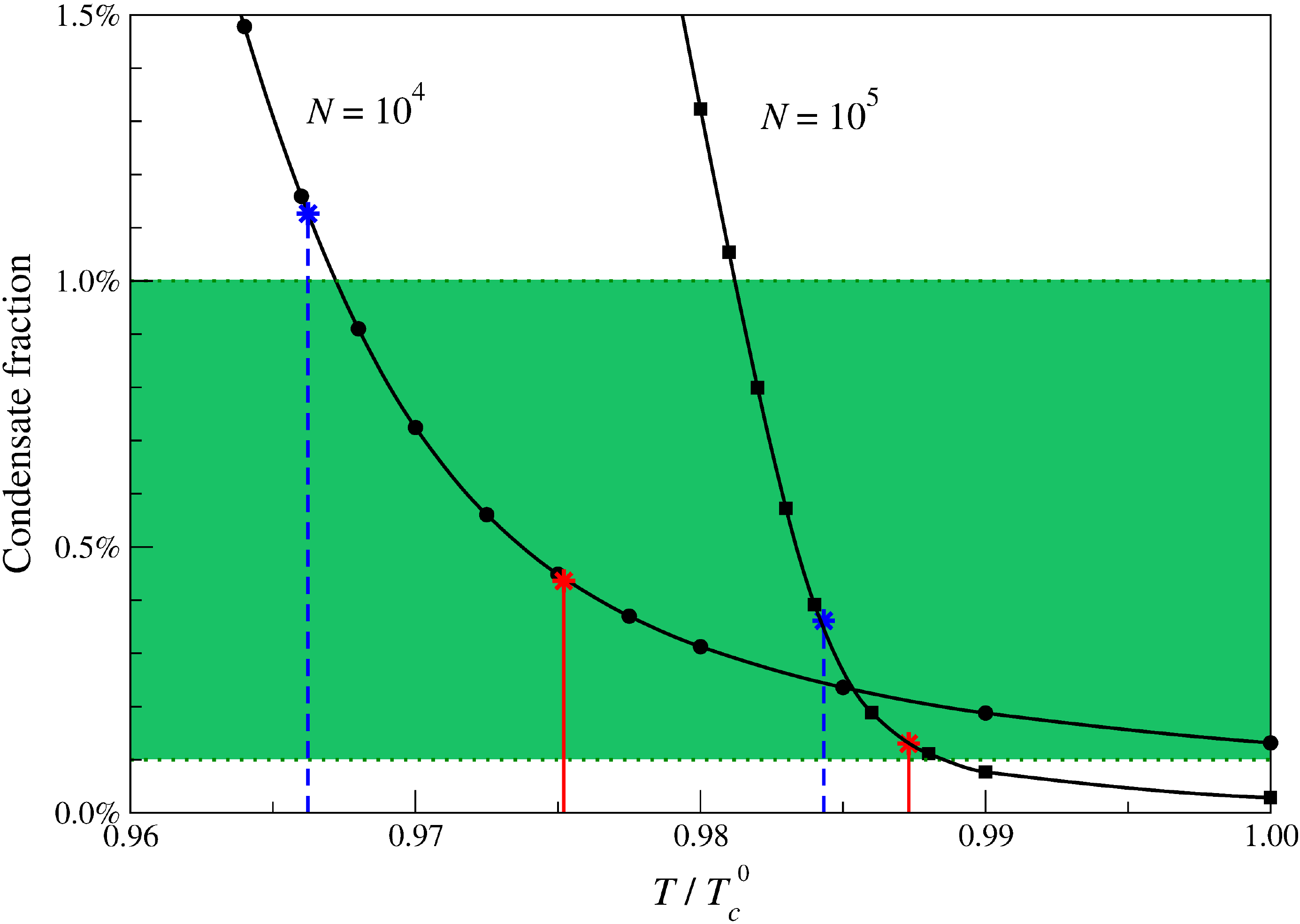}
\caption{\label{fig2}(Color online) Condensate fraction as a function of the rescaled temperature $T/T_c^0$. The curve seen on the left with solid circles is for $N = 10^4$ and the curve on the right with solid squares is for $N = 10^5$. The condensation temperatures calculated at first order and with our generalized formula are shown as dashed blue and solid red lines respectively. The green area represents a typical range where the condensation temperature can be detected experimentally\,\cite{PRL2004,PRL2011}.}
\end{center}
\end{figure}

Let us now look at the results presented by Noronha in more details. In the case of an isotropic trap, we reproduce in Fig.\,\ref{fig1} the variation of the rescaled condensation temperature $T_c/T_c^0$  as a function of $\log_{10}(N)$ in the range of validity of our model, \textit{i.e.} for $N \geqslant 10^4$. The dashed blue line shows the condensation temperature calculated with the well known first-order finite size correction, while the solid red line shows the result of our generalized LDA formula\,\cite{PRA2011}, which includes higher-order corrections. The temperatures $T_{0.1\%}$, $T_{0.5\%}$ and $T_{1\%}$ at which the condensate fraction reaches 0.1\pct{}, 0.5\pct{} and 1\pct{} are shown as different dotted green lines. These three additional temperatures have been evaluated directly from thermodynamic sums over the discrete energy levels of the trap. We see in Fig.\,\ref{fig1} that the first order formula lies between $T_{1\%}$ and $T_{0.1\%}$, while our result lies between $T_{0.5\%}$ and $T_{0.1\%}$. From this graph, one can only conclude that the two analytical approximations give similar and realistic results, which converge to the same thermodynamic limit for large $N$ values.

In addition, in Fig.\,\ref{fig2} we show the condensate fraction as a function of the rescaled temperature $T/T_c^0$. The curve seen on the left with solid circles is for a number of atoms $N = 10^4$ and the curve on the right with solid squares is for $N =10^5$. The condensation temperatures calculated with the first order correction formula and with our generalized LDA formula\,\cite{PRA2011} are shown as vertical dashed blue and solid red lines respectively, using the same color code as in Fig.\,\ref{fig1}. We see here, again, that our generalized formula yields a slightly higher condensation temperature than the one obtained with the first order approximation. It is therefore associated with a smaller condensate fraction.

In typical experiments, the condensation temperature is measured by fitting the bi-modal distribution of the gas. Using such a scheme, condensate fractions as low as 0.1\pct{}--1\pct{} can be detected\,\cite{PRL2004,PRL2011}. It is from such a detection that the value of $T_c$ can be inferred experimentally. This typical range of detection (0.1\pct{}--1\pct{}) is represented by the green area of Fig.\,\ref{fig2}. We see here that the condensation temperatures calculated at first order and with our approach\,\cite{PRA2011} (presented as star symbols in Fig.\,\ref{fig2}) are both located in this windows or in its immediate vicinity. Our generalized formula (solid red line with stars in Fig.\,\ref{fig2}) yields a condensation temperature which defines rather well the very beginning of the condensation process, when the condensate fraction does not yet vary too much with the temperature. It is fully compatible with typical experimental determinations of $T_c$.

Finally, in the last part of the comment, J. Noronha discusses different results of the literature obtained with power-law traps higher than cubic without settling the conflict which exists between these different results in these particular types of traps. We note that such a debate has also existed for several decades concerning the influence of atomic correlations on the condensation temperature\,\cite{RMP1999} before this question was finally settled in 1999\,\cite{PRL1999}. Answering this question for finite size effects requires additional investigations.

\bibliographystyle{apsrev4-1}

\end{document}